\documentclass[prl,twocolumn,superscriptaddress,showpacs]{revtex4}

\usepackage{graphicx}

\newcommand{\etal}{{\em et al}.}

\newcommand{\EF}{$E_{\rm F}$}

\newcommand{\vo}{V$_2$O$_3$}

\newcommand{\vcro}{(V$_{1-x}$Cr$_x$)$_2$O$_3$}

\begin{document}

\title{Prominent quasi-particle peak in the photoemission spectrum \\ of the metallic phase of \vo}

\author{S.-K. Mo}
\affiliation{Randall Laboratory of Physics, University of
Michigan, Ann Arbor, MI 48109}
\author{J. D. Denlinger}
\affiliation{Advanced Light Source, Lawrence Berkeley National
Laboratory, Berkeley, CA 94720}
\author{H.-D. Kim}
\affiliation{Pohang Accelerator Laboratory, Pohang 790-784, Korea}
\author{J.-H. Park}
\affiliation{Department of Physics, Pohang University of Science
and Technology, Pohang 790-784, Korea}
\author{J. W. Allen}
\affiliation{Randall Laboratory of Physics, University of
Michigan, Ann Arbor, MI 48109}
\author{A. Sekiyama}
\author{A. Yamasaki}
\author{K. Kadono}
\author{S. Suga}
\affiliation{Department of Material Physics, Graduate School of
Engineering Science, Osaka University, 1-3 Machikaneyama,
Toyonaka, Osaka 560-8531, Japan}
\author{Y. Saitoh}
\affiliation{Department of Synchrotron Radiation Research, Japan
Atomic Energy Research Institute, SPring-8, Sayo, Hyogo 679-5143,
Japan}
\author{T. Muro}
\affiliation{Japan Synchrotron Radiation Research Institute,
SPring-8, Sayo, Hyogo 679-5143, Japan}
\author{P. Metcalf}
\affiliation{Department of Physics, Purdue University, West
Lafayette, IN 47907}
\author{G. Keller}
\affiliation{Theoretical Physics III, Center for
Electronic Correlations and Magnetism, University of Augsburg,
86135 Augsburg, Germany}
\author{K. Held}
\affiliation{Max Planck Institute for Solid State Research,
Heisenbergstrasse 1, D-70569 Stuttgart, Germany}
\author{V. Eyert}
\affiliation{Theoretical Physics II, University of Augsburg,
86135 Augsburg, Germany}
\author{V.\ I.\ Anisimov}
\affiliation{Institute of Metal Physics, Ekaterinburg GSP-170,
Russia}
\author{D. Vollhardt}
\affiliation{Theoretical Physics III, Center for
Electronic Correlations and Magnetism, University of Augsburg,
86135 Augsburg, Germany}

\date{Received \hspace*{30mm}}

\begin{abstract}
We present the first observation of a prominent quasi-particle
peak in the photoemission spectrum of the metallic phase of \vo\
and report new spectral calculations that combine the local
density approximation with the dynamical mean-field theory (using
quantum Monte Carlo simulations) to show the development of such a
distinct peak with decreasing temperature. The experimental peak
width and weight are significantly larger than in the theory.
\end{abstract}

\pacs{PACS numbers: 71.20.Be, 71.30.+h, 79.60.-i}

\maketitle

%   Introduction

\vcro\ displays a complex phase diagram with paramagnetic metal
(PM), paramagnetic insulator (PI) and antiferromagnetic insulator
(AFI) regions.  The PM to PI transition serves as the paradigm of
the Mott-Hubbard (MH) metal-insulator transition (MIT)
\cite{Mott}. The MH scenario for \vcro\ was put forth originally
in the context of the half-filled one-band Hubbard model in which
the tendency of the on-site Coulomb repulsion `$U$' to make a
correlation gap insulator competes with the tendency of site to
site hopping to make a broad band metal of bandwidth `$B$'. A
coherent thermodynamically consistent description of the MIT
became possible with the development of the dynamical mean-field
theory (DMFT) \cite{DMFT}.  DMFT describes the strongly
interacting metal in terms of Fermi liquid quasi-particles, i.e.
single particle excitations near the Fermi energy \EF\ which
remain well defined as in a non-interacting system but have a self
energy correction that increases their effective mass and reduces
their spectral weight. In application to the Hubbard model, DMFT
is significant as the best description that can be made by using a
local (i.e. independent of momentum {\bf k}) self energy.  It may
be formulated as a mapping of the lattice problem onto an
effective Anderson impurity model coupled self-consistently to an
effective conduction band bath \cite{DMFTmapping}.   In the
metallic phase, although the large $U$ value acts to separate much
of the band's spectral weight away from \EF\ into the so called
upper and lower Hubbard bands, there remains at \EF\ a distinctive
quasi-particle (QP) peak, not accidentally reminiscent of the
Kondo/Suhl-Abrikosov resonance \cite{KSAtheory} of the Anderson
impurity model.  The weight of the QP-peak decreases with
increasing $U/B$ and goes to zero at a critical value of $U/B$,
which thus marks the MIT.

Such a distinctive peak could in principle be seen in
photoemission spectroscopy (PES).  However, in spite of continuing
efforts for over twenty years, literature \cite{Sawatzky, Smith88,
Smith94, Shin, Kim, Ralph} V $3d$ PES spectra for the PM phase of
\vcro\ have shown at most a near \EF\ feature that is the smallest
part of the spectrum. One could hypothesize that the distinctive
central peak of the half-filled one-band model is obscured and
washed out by the multi-band complexity of the actual electronic
structure of \vcro\, in which the two 3$d$-electrons of the
V$^{3+}$ ion must be distributed among singly degenerate $a_{1g}$
and doubly degenerate $e_g^\pi$ orbitals derived from a small
trigonal crystal field splitting of the cubic $t_{2g}$ manifold of
the V $3d$ states.   Indeed, recent experiment \cite{Parkprb} and
theory \cite{Ezhov} establishing an S=1 state for the V$^{3+}$
ions show that such a multi-band description is essential.
Further, a recent study that included these material specific
aspects by combining the local density approximations with DMFT
(using quantum Monte Carlo simulations) [LDA+DMFT (QMC)]
\cite{Held} found for the PM phase a weak \EF\ peak in good
agreement with that of a well resolved and high quality PM phase
spectrum by Schramme \etal \cite{Schramme}, taken at a photon
energy ($h\nu$) of $60$~eV. However, it is important to note that
numerical complexities constrained the calculation to be performed
at a temperature $T$ of $1160$~K, whereas the photoemission
spectrum was obtained at a much lower temperature of $300$~K, and
that for the photon energy used, recent PES studies of other
vanadium oxides have been found \cite{Maiti, sekiyama2} to yield
spectra not characteristic of the bulk, but rather of the surface
atoms whose lower coordination number reduces $B$ proportionately
and thus can render the surface layer much more strongly
correlated or even insulating.  Both circumstances go in the
direction of a greatly reduced QP peak.

%Inserting Figure 1
\begin{figure}[!t]
\includegraphics[width=2.9 in]{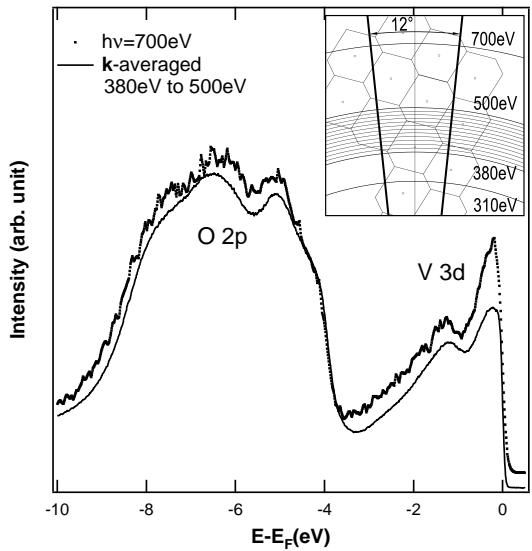}
\caption{\label{Fig 1} $h\nu=700eV$ and {\bf k}-averaged
 PES spectra of \vo. The inset shows the {\bf k}-space covered
by the detector angular acceptance for various $h\nu$.}
\end{figure}

In this Letter we report angle-integrated PES spectra for the
metallic phase of \vo\ that are different from all previous
spectra in showing an \EF\ peak whose amplitude exceeds that of
the rest of the V $3d$ spectrum by a factor approaching two. These
results were achieved by measuring at high photon energies in order
to increase the bulk sensitivity but with much better resolution
\cite{Sekiyama} than in previous \cite{Park, Ralph} studies at
these energies.  Equally important is the use of a small photon
spot ($\approx 100\mu$m diameter) that we have found \cite{smallspot}
 to be essential for minimizing the effects of steps and other roughness
that can be seen in high resolution optical micrographs of typical
cleaved single crystal surfaces of this material.  So doing
is very important because atoms with the edge and corner (i.e. acute)
geometries of steps and general roughness have coordinations
that are reduced from those of the planar surface atoms.
Concomitantly, as is known from past theory \cite{Johansson}
 and experiment \cite{Kaindl} on correlated electron materials,
they have even larger differences from the electronic
structure of the bulk than occurs for the planar surface atoms.
The large amplitude of the newly
observed peak disagrees with the theory of Ref. \cite{Held}.
However, we also report new DMFT+LDA(QMC) calculations for the
same electronic structure and $U$ value as previously \cite{Held},
showing that with temperature decreasing to $300$~K such a
prominent central peak develops.  Nonetheless the quantitative
comparison enabled by the new experiment and new theory reveals
important differences.

%   Experiment

PES using circularly polarized photons with $h\nu$ between
$310$~eV and $700$~eV was performed at the twin-helical undulator
beam line BL25SU \cite{BL25SU} of SPring-8 equipped with a SCIENTA
SES200 analyzer and giving a photon spot size of diameter $100
\mu$m. The Fermi level and overall energy resolution ($\approx
90$~meV to $170$~meV over the $h\nu$ range) were determined from
the Fermi-edge spectrum of a Pd metal reference. Well annealed
oriented single-crystalline samples of \vo\ were cleaved to expose
a hexagonal (10$\bar{1}$2) plane in a vacuum of $2 \times
10^{-10}$~Torr.  Using a closed-cycle He cryostat and an embedded
resistive heater, the sample temperature was held at $T = 175$~K,
somewhat above the PM/AFI transition. Surface integrity was well
maintained under the vacuum and photon exposure, as shown by the
repeatability of the spectra for at least ten hours after
cleaving, after which the PM/AFI transition could still be
observed in the spectra at the temperature appropriate for
stoichiometric \vo. The inset to Fig. 1 shows a cross-section of
the {\bf k}-space Brillouin zone stacking  normal to the cleavage
plane and also the spherical surfaces \cite{inner} traversed as
the detector angles are varied about the normal for various fixed
$h\nu$ between $310$~eV and $700$~V.  Radial lines show the {\bf
k}-range corresponding to the analyzer acceptance angle of about
$\pm 6$~degrees. This range is $\pm 1.2$~\AA$^{-1}$ for a
photoelectron with $500$~eV kinetic energy and covers more than
one Brillouin zone for any $h\nu$ arc.

%   Figure 1

The $h\nu$ = $700$~eV spectrum of Fig.\ 1 shows the general
character of the data.  The V $3d$ emission is well separated from
the O $2p$ emission, and the newly observed prominent \EF\
peak is very obvious. We are interested here in the $h\nu$
dependence of the spectral shape of the V $3d$ emission.
Although there is a
huge $3d$ enhancement over the range $510$~eV to $560$~eV,
due to a $3d$ cross-section resonance at the V $2p$ edge
\cite{Park, Ralph}, we have deliberately
avoided the resonance region because we have found
\cite{Auger} in this range incoherent Auger
emission that distorts the spectral shape.  Because we make
comparisons to a {\bf k}-summed theory, Fig. 1 also shows the
result of {\bf k}-averaging across a full Brillouin zone by
using spectra taken in $10$~eV steps from $380$~eV to
$500$~eV.  The \EF\ peak in the {\bf k}-averaged spectrum
is somewhat smaller than that in the $700$~eV
spectrum.  This is however due to the reduced bulk sensitivity
of the lower photon energies rather than the {\bf k}-selectivity
of the single photon energy, as discussed next.

%Inserting Figure 2
\begin{figure}[!b]
\includegraphics[width=2.9 in]{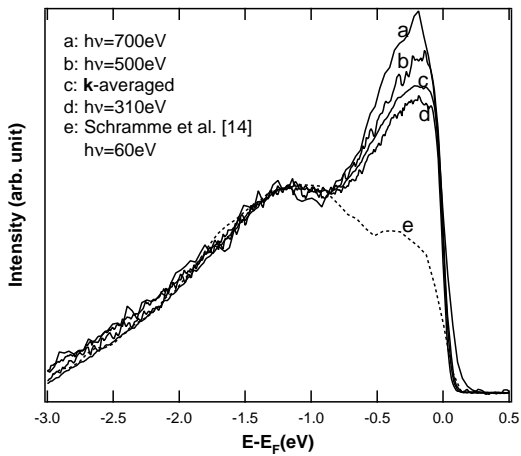}
\caption{\label{Fig 2} PES spectra taken with various $h\nu$, the
largest of which yields the greatest bulk sensitivity.}
\end{figure}

%   Figure 2

Fig.\ 2 shows our V $3d$ spectra for $h\nu$ from $310$~eV to
$700$~eV along with the $h\nu$ = $60$~eV spectrum \cite{Schramme}
that was found to compare favorably to the $1160$~K
theory spectrum in Ref. \cite{Held}.
A Shirley-type inelastic background has been removed in an
identical way \cite{Shirley} for each spectrum. The spectra
are normalized over the range below $- 1$~eV
for ease of comparing their \EF\ peaks. Relative to the
spectrum below $- 1$~eV, the \EF\ peaks in the spectra of Fig.\ 2
increase monotonically with increasing $h\nu$.  Over the range
for the {\bf k}-averaged spectrum the peak is never larger than
at $500$~eV.  Although some non-monotonic
variation over this range and also below $300$~eV
(not shown) could signal {\bf k}-dependence, we conclude
that above $300$~eV the change of probe depth dominates
{\bf k}-dependent effects in the data. This conclusion is reinforced
by the observation at $500$~eV that rotating the
detection angle away from the normal in steps of $15$~degrees out to
$60$~degrees, which monotonically decreases the effective probe depth,
monotonically decreases the \EF\ peak to be comparable to
the \EF\ feature that is seen in the $60$~eV data.
Reliably estimating a bulk spectrum from the $h\nu$ and angle
dependences using various phenomenological models
is not simple.  We will present such estimates in a more
detailed paper.  However, the results do not differ from
the 700 eV spectrum itself in any way that is
significant for the comparison to new theory that we make next.

%Inserting Figure 3
\begin{figure}[!b]
\includegraphics[width=2.9 in]{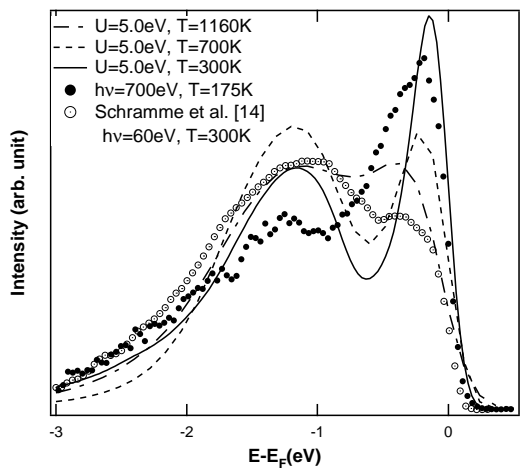}
\caption{\label{Fig 3} LDA+DMFT(QMC) calculations made at three
temperatures for $U=5.0$~eV, the $h\nu$ = $60$~eV PES spectrum
\cite{Schramme} compared to theory in Ref. \cite{Held}, and our
new $h\nu$ = $700$~eV bulk sensitive spectrum.}
\end{figure}

%Theory

On the theoretical side, new, very CPU-intensive LDA+DMFT (QMC)
calculations have been made at considerably lower temperatures
than previously \cite {Held}, i.e. $T = 700$~K and $T = 300$~K, in
exactly the same manner \cite {LDADMFT} as before.  The LDA
density of states of the three $t_{2g}$ bands at the Fermi energy
was identified with a one-particle Hamiltonian which was
supplemented  by a local Coulomb interaction  $U\!=\!5\,$eV (only
$U$ values close to $5\,$eV generate the experimentally observed
MIT upon Cr-doping \cite{Held}) and a Hund's rule coupling
$J=0.93\,$eV (according to constrained LDA calculations
\cite{Solovyev96}). This three-band many-body problem was solved
by DMFT using QMC simulations.

%   Figure 3

Fig.\ 3 shows our theory spectra for three temperatures, compared
to the experimental $h\nu$ = $60$~eV spectrum ($300$~K) and the
$700$~eV spectrum ($175$~K) of Fig.\ 2, all scaled to have equal
areas. The theory curves include the Fermi function for the
appropriate temperature and Gaussian broadening of 90 meV to
simulate experimental resolution. Qualitatively similar to the
behavior of the impurity model's Kondo resonance \cite{KSAtheory},
at higher temperatures the QP peak is greatly broadened by thermal
occupation of states bearing local magnetic moments. As the system
settles with decreasing temperature toward its non-magnetic ground
state, the peak sharpens, its amplitude increases strongly, and it
becomes clearly separated from the lower Hubbard band at about
$-1.25$eV.  These temperature-induced changes strongly enhance the
differences between theory and experiment that were already
present but small in Ref. \cite{Held}, i.e., the amplitude of the
QP peak is much larger in the theory than in the $h\nu$ = $60$~eV
experimental spectrum. Qualitatively the comparison to the new
bulk sensitive spectrum is much better, with the prominent QP peak
and the lower Hubbard band of the theory being very similar,
respectively, to the newly observed \EF\ peak and the broad hump
centered at $- 1.25$~eV.  Strikingly different, however, are that
the experimental \EF\ peak is broader and also has more spectral
weight.  These differences will persist for theory at the slightly
lower temperature of the data.  The width difference could
increase somewhat.  The weight difference would decrease but
probably not greatly.  From $1160$~K to $300$~K the theory peak's
integrated spectral weight, taken as that above $-0.63$~eV,
increases by only 11$\%$.

Within the DMFT scheme the larger experimental
width and weight implies weaker correlation that could be
described by using a reduced $U$-value.  However in the
present theory \cite {Held} the MIT under
Cr-doping would not occur for such a reduced $U$-value
unless $U$ then increases with Cr doping or through
the MIT transition.  Either might be expected to occur
through reduction in various effective screening processes, some
of which are excluded from the present theory by its
restriction to three bands.  Another possible origin
for the increased width of the experimental peak is
the {\bf k}-dependence of the single particle self energy.
This is beyond our current calculations, but may partly be
included by an improved DMFT calculation which uses a pair
of V ions \cite {pairs} instead of a single site.

%   Summary

In summary we have discovered a prominent \EF\
peak in the PES spectrum of \vo\ in the PM phase,
at a temperature somewhat above that of the AFI
phase transition.  New
LDA+DMFT(QMC) theory for a comparable temperature also
finds such a prominent QP peak, generic to the DMFT
theory of the Hubbard model near the MIT transition.
The quantitative comparison that is enabled by the
two new results reveals significant
differences that must be addressed in future work.
Nonetheless, considering the various approximations of the present
theory, the comparison is very encouraging as to the occurrence
of the basic QP peak that is central to the DMFT description.

%\acknowledgements
\begin{acknowledgments}
This work was supported by the U.S. NSF at the University of
Michigan (UM) (Grant No.~DMR-99-71611), by the U.S. DoE at UM
(Contract No.~DE-FG-02-90ER45416) and at the ALS (Contract No.~
DE-AC03-76SF00098), by Grant-in-Aid for COE Research (10CE2004) of
MEXT, Japan by JASRI (No.~2000B0335-NS-np), by KOSEF through eSSC
at POSTECH, by the Deutsche Forschungsgemeinschaft (DFG) through
SFB 484, by the Emmy-Noether program of the DFG, by the Russian
Foundation for Basic Research grant RFFI-01-17063, by the
Alexander von Humboldt-Foundation, and by the
Leibniz-Rechenzentrum, M\"{u}nchen.
\end{acknowledgments}


\begin{thebibliography}{99}
\bibitem{inner} Following Ref. \cite{Smith88} the figure is drawn for
an inner potential of $21.76$~eV, but in fact there is negligible
dependence on the inner potential value for these large $h\nu$.
\end{thebibliography}
\end{document}